# Should Students be Provided Diagrams or Asked to Draw Them While Solving Introductory Physics Problems?

Alex Maries and Chandralekha Singh

*Department of Physics and Astronomy, University of Pittsburgh, Pittsburgh, PA 15260*

**Abstract.** Drawing appropriate diagrams is a useful problem solving heuristic that can transform a given problem into a representation that is easier to exploit for solving it. A major focus while helping introductory physics students learn problem solving is to help them appreciate that drawing diagrams facilitates problem solution. We conducted an investigation in which 111 students in an algebra-based introductory physics course were subjected to two different interventions during recitation quizzes throughout the semester. They were either (1) asked to solve problems in which the diagrams were drawn for them or (2) explicitly told to draw a diagram. A comparison group was not given any instruction regarding diagrams. We developed a rubric to score the problem-solving performance of students in different intervention groups. Here, we present some surprising results for problems which involve considerations of initial and final conditions.

## INTRODUCTION

Drawing diagrams is a useful problem solving heuristic whose importance cannot be over-emphasized. Diagrammatic representations have been shown to be superior to verbal representations when solving problems [1-3]. This is one reason why physics experts automatically employ diagrams in attempting to solve problems. However, introductory physics students need explicit help understanding that drawing a diagram is an important step in organizing and simplifying the given information into a representation which is more suitable to mathematical manipulation. Previous research shows that students who draw diagrams even if they are not rewarded for it are more successful problem solvers [4]. We extend that research here and investigate how the student performance will be affected when students are given a diagram instead of being asked to draw it. Here, we focus on some surprising results for two problems involving considerations of initial and final conditions.

## METHODOLOGY

A class of 111 algebra-based introductory physics students was broken up into three different recitations. All recitations were taught in the traditional way in which the TA worked out problems similar to the homework problems and gave a 15 minute quiz at the end. Students in all recitations attended the same lectures, were assigned the same homework, and had the same exams and quizzes. In the recitation quizzes throughout the semester, the three groups were given the same problems but with the following interventions: in each quiz problem, in addition to the problem statement, the first intervention group (Group 1) was given specific instructions to draw a diagram; the second intervention group (Group 2) was given a diagram drawn by the instructor that was meant to aid in solving the problem and the third group (Group 3) was the comparison group and was not given any diagram nor an explicit instruction to draw one.

The sizes of the different recitation groups varied from 22 to 55 students because the students were not assigned a particular recitation, they could go to whichever recitation they wanted. For the same reason, the sizes of each recitation group also varied from week to week, although not as drastically because most students ($\approx 80\%$) would stick with a particular recitation. Furthermore, each intervention was not matched to a particular recitation. For example, in one week, students in the Tuesday recitation comprised the comparison group, while another week the comparison group was a different recitation section. This is important because it implies that individual students were subjected to different interventions from week to week and we do not expect cumulative effects due to the same group of students always being subjected to the same intervention.

In order to ensure homogeneity of grading, we developed rubrics for each problem we analyzed and made sure that there was at least 90% inter-rater-reliability between two different raters. The rubrics were developed through an iterative process. During the development of the rubric, the two graders also discussed a student's score separately from the one obtained using the rubric and adjusted the rubric if it was agreed that the version of the rubric was too stringent or too generous. After each adjustment, all students were graded again on the improved rubric.

For the two problems analyzed in this paper, our goal was to investigate if there were any statistical differences in the scores of the groups of students subjected to different interventions.

The two problems discussed and the diagrams given to students in Group 2 are the following:

**Problem 1**

"Two identical point charges are initially fixed to diagonally opposite corners of a square that is 1 m on a side. Each of the two charges $q$ is 3 C. How much work is done by the electric force if one of the charges is moved from its initial position to an empty corner of the square?"

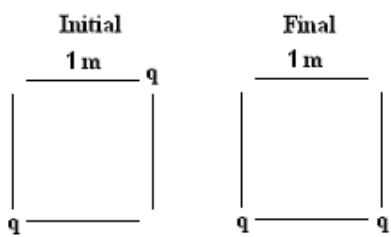

**FIGURE 1.** Diagram for problem 1 given to students in Group 2.

**Problem 2**

"A particle with a mass $10^{-5}$ kg and a positive charge of 3 C is released from rest from point A in a uniform electric field. When the particle arrives at point B, its electrical potential is 25 V lower than the potential at A. Assuming the only force acting on the particle is the electrostatic force, find the speed of the particle when it arrives at point B."

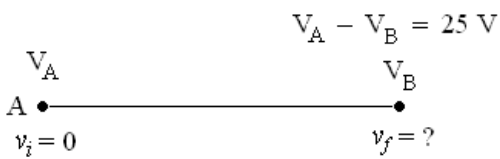

**FIGURE 2.** Diagram for problem 2 given to students in Group 2.

These diagrams were drawn by the instructor and they are very similar to what most physics experts would generally draw in order to solve the problem. Furthermore, the second diagram also includes an important piece of information from the problem that would normally be included in a known quantities/target quantities section of a solution. Neither diagram was meant to trick the students, but rather they were provided as possible scaffolding support for them.

As mentioned earlier we developed rubrics for each problem. In Table 1, we provide the summary of the rubric for the first problem. The rubric for the other problem was developed in a similar manner.

**TABLE 1.** Summary of the rubric for problem 1.

| Ideal Knowledge | |
|---|---|
| 1. W = -qΔV or W = -ΔEPE | 2p |
| 2. Getting $V_f$, $V_i$ and doing ΔV = $V_f$ - $V_i$ **or** $EPE_f$, $EPE_i$ and doing ΔEPE = $EPE_f$ –$EPE_i$ | 7p |
| 3. Units | 1p |
| **Incorrect Ideas** | |
| Used the electrostatic force incorrectly: if provided units (-8 p), no units (-10 p) | |
| 1. Used incorrect equation (-2 p) | |
| 2.1 Got one potential or one EPE wrong (-2 p) | |
| 2.1 Got both potentials or EPEs wrong (-4 p) | |
| 2.2 Did not subtract (-2 p), and/or other mistakes (-3 p / -1 p) | |
| 2.3. Wrong sign (-1 p) | |
| 3. Wrong or no units (-1 p) | |

Table 1 shows that there are two parts to the rubric: Ideal Knowledge and Incorrect Ideas. Table 1 also shows that in the Ideal Knowledge part, the problem was divided into different sections and points were assigned to each section (10 maximum points). Each student starts out with 10 points and in the Incorrect Ideas part we list the common mistakes students made and how many points we deducted for each of those mistakes. Using the electrostatic force for this problem is not an effective strategy for algebra based students (this approach involves calculus), so students who attempted to use force had to be graded separately because their approach is not productive. The rest of the rubric in the "Incorrect Ideas" part was used for grading the students with a productive approach. For each mistake, we deduct a certain number of points. We note that it is not possible to deduct more points than a section has (e.g., the two mistakes that are both labeled 2.1 in Table 1 are mutually exclusive). We also left ourselves a small window (labeled 2.2) if the mistake a student made was not in the rubric.

## RESULTS

Before discussing the findings for the two problems outlined, we note that this investigation was carried out throughout an entire semester and we analyzed the student performance in the two interventions and the comparison group on more than 10 problems. For a majority of problems there were no significant differences between the scores of the different intervention groups and comparison group. Also, students in the intervention groups never performed better than those in the comparison group. Furthermore, the two problems analyzed here are the only ones that involved relating initial and final conditions and were part of the same three problem recitation quiz. In the third problem of that quiz, we did not find any statistical differences between the different groups. We therefore believe that the groups are not different from each other in terms of students' ability and any differences in student performance on these problems are due to the interventions.

**TABLE 2**. Group sizes (N), averages, and standard deviations on the two problems.

| Problem 1 | N  | Average | Std. dev. |
|-----------|----|---------|-----------|
| Group 1   | 26 | 8.5     | 1.88      |
| Group 2   | 34 | 6.9     | 2.82      |
| Group 3   | 51 | 9.0     | 1.39      |
| **Problem 2** | N  | Average | Std. dev. |
| Group 1   | 26 | 9.0     | 1.44      |
| Group 2   | 34 | 6.4     | 3.06      |
| Group 3   | 51 | 8.6     | 1.34      |

For both problems, Group 2 was the one given the diagram with the problem statement. It is evident from Table 2 that students who were given the diagram performed significantly worse than all the others; their averages are lower by roughly 20% compared to the other groups. We also performed an analysis of variance (ANOVA) with $t$-tests. The p-values are shown in Table 3.

**TABLE 3.** p values for ANOVA between the different groups.

|           | Groups 1-2 | Groups 2-3 | Groups 1-3 |
|-----------|------------|------------|------------|
| **Problem 1** | 0.015   | < 0.001    | 0.193      |
| **Problem 2** | < 0.001 | < 0.001    | 0.342      |

Table 3 shows that Group 2 (which was given the diagram) did significantly worse than the other two groups. More noteworthy is how small the p values are (three of them being less than 0.001). Table 3 also shows that the scores of Groups 1 and 3 are comparable. We note that, for problem 1, virtually all students drew a diagram even if they were not specifically asked to do so. However, for problem 2, only 57% of the students in Group 3 drew a diagram. Even though the students from Group 3 who drew diagrams performed better than those who did not, we did not find a statistically significant difference between their scores. We performed a *t*-test to compare the performance of students in Group 3 who did not draw a diagram and all students in Group 2. We found that Group 2 did significantly worse (p = 0.004). Thus, on problem 2, students who did not draw a diagram did better than those who were given a diagram drawn by the instructor. Some possible reasons for this counter-intuitive result will be discussed later.

**TABLE 4.** Percentages (and numbers) of students in each group that obtained above 8 or 5, 6 and 7 or below 5 (out of 10).

| Problem 1 | ≥8       | 5,6,7    | ≤4       |
|-----------|----------|----------|----------|
| Group 1   | 73% (19) | 23% (6)  | 4% (1)   |
| Group 2   | 53% (18) | 21% (7)  | 26% (9)  |
| Group 3   | 82% (42) | 16% (8)  | 2% (1)   |
| **Problem 2** | ≥8   | 5,6,7    | ≤4       |
| Group 1   | 81% (24) | 15% (4)  | 4% (1)   |
| Group 2   | 41% (17) | 21% (7)  | 38% (13) |
| Group 3   | 76% (47) | 22% (11) | 2% (1)   |

Table 4 shows that the percentage of students who performed poorly on this problem (obtained a score less than 5 out of 10) from Group 2 is significantly larger than those in Groups 1 and 3 but percentages with an intermediate score are comparable.

## DISCUSSION

Prior research has shown that students in classes that emphasize conceptual learning and employ active-learning methods outperform students from traditional classes even on quantitative tests [5]. It is possible that students who perform poorly on physics problem solving may do so not because they have poor mathematical skills, but rather because they do not effectively analyze the problem conceptually. In particular, they may not employ effective problem solving heuristics and transform the problem into a representation which makes further decision making easier. For example, converting a physics problem from the verbal to the diagrammatic representation by drawing a diagram is a heuristic that can facilitate better understanding of the problem and aid in solving it.

One hypothesis for why students in Group 2 who were given a diagram performed significantly worse than the other two groups is that they were more likely to skip the important step of conceptual analysis of the problem because the diagram was already provided. Therefore, they had difficulty in conceptualizing the problem and formulating a correct solution. The data in Table 4 suggests that students in Group 2 on average did significantly worse because more students in that group than in the other groups did very poorly. The fact that many students who were given the diagrams failed to understand the problem conceptually is also evident from observing their individual solution strategies. For example, more students in Group 2 than in the other groups had a formula-based approach and it was unclear by observing their written work how they arrived at the decision to use those formulas (which were often not productive).

As mentioned earlier, in problem 2, even the students who did not draw a diagram from the comparison group did better than the students in Group 2. One possible reason may be that problem 2 (actually, both problems discussed here) is not a difficult or multi-part problem requiring the use of many physics principles. Therefore, the cognitive load while solving the problem may not be very high even if an explicit diagram is not drawn and students keep the relevant information in their working memory while solving the problem. Students' written work from the three groups also suggests that a higher percentage of students who are not given the diagram were going through an explicit process of making sense of the problem than those students who were given the diagram.

As mentioned earlier, students in Group 2 did not do worse than those in the other groups on a majority of other problems not discussed here. What makes these problems special in this respect is not obvious. However, we note that what these two problems have in common other than the fact that they are both from electrostatics is that they both involve considerations of initial and final conditions. This latter characteristic was not present in any of the other problems we analyzed.

To evaluate the opinions of other instructors who had taught introductory physics frequently, we presented the three interventions for the two problems discussed here to seven physics faculty members and asked them to predict which group is likely to perform the best. Interestingly, some faculty members automatically assumed that the diagram would help and tried to answer the question "why would the diagrams help students" despite the fact that we asked them a neutral question about the group which is likely to perform the best. Also, similar to our original hypothesis, all seven faculty members incorrectly predicted that students in Group 2 would perform the best because they were given explicit diagrams clarifying the situation. Some of them also mentioned that the second problem discussed here is more difficult than the first and that the given diagram should help more with the first problem than the second one because the first problem involves a situation with charges situated in two-dimensions.

When the faculty members were told how the students actually performed, two of them recalled that they have observed in the past that giving a diagram has sometimes worsened student performance. Some of them mentioned that when they themselves solve a physics problem, they perform an initial conceptual analysis and often draw a diagram to make the situation clearer. Similar to our hypothesis, they noted that the absence of this important stage of problem solving when a diagram is provided to students can derail the entire problem solving process. Others noted that when a diagram is given, students may not read the problem statement carefully. Some claimed that for the first problem, students in Group 2 were more likely to resort to a solution method involving force instead of energy because students are more likely to encounter diagrams with charges at the corner of a square or rectangle in problems involving the electrostatic force in books and homework problems.

When the faculty members were explicitly asked whether their students would find any aspect of the diagrams confusing, their responses were negative. The disconnect between the faculty members' initial predictions about the usefulness of providing diagram and students' actual performance further suggests that the manner in which the cognitive processes of the novices was negatively affected by the given diagrams is quite complex. In the future, we plan to conduct think-aloud interviews with students while they solve these two problems with or without the diagrams. This may help identify the missing cognitive processes for those who were given the diagram with the problems and shed light on why they are likely to perform worse.

## CONCLUSIONS

We found that students who were provided diagrams for the two problems discussed here performed significantly worse than the ones who were not given diagrams. In the future, we will use think-aloud interviews to explore how providing a diagram may sometimes inhibit cognitive processes that favor understanding and encourage students to make simplistic attempts at a solution (often formula-seeking in nature) as observed in this study.